\documentclass[11pt]{article}
\usepackage{setspace}
\usepackage{blindtext, graphicx}
\usepackage{graphicx}
\usepackage{multirow}
\usepackage{floatrow}
\usepackage[T1]{fontenc}
\usepackage{capt-of}
\DeclareFloatFont{tiny}{\scriptsize}%
\usepackage[utf8]{inputenc}
\usepackage{dtklogos} 
\usepackage{subfigure}

\usepackage{subfig} 
\usepackage{caption}

\usepackage{mathtools}
\usepackage{listings}
\usepackage{color}
\usepackage{soul}
\usepackage{array}
\newcolumntype{P}[1]{>{\centering\arraybackslash}p{#1}}
\usepackage{setspace} 
\usepackage{graphicx}
\usepackage{algcompatible}
\usepackage{afterpage,lipsum}
\usepackage{tikz}
\usepackage{longtable}
\usepackage{multirow}
\usepackage{pbox}
\usepackage{url}
\usepackage{lscape}
\usepackage{verbatim} 
\usepackage{array}
\usepackage{booktabs}
\usepackage{adjustbox,lipsum}
\usepackage[ruled,vlined,linesnumbered]{algorithm2e}
\usepackage{amsmath}
\usepackage{amsmath}
\usepackage{mathtools}
\usepackage{anyfontsize}

\usepackage{multicol}
\newcommand{\specialcell}[2][c]{%
  \begin{tabular}[#1]{@{}c@{}}#2\end{tabular}}
\definecolor{dkgreen}{rgb}{0,0.6,0}
\definecolor{gray}{rgb}{0.5,0.5,0.5}
\definecolor{mauve}{rgb}{0.58,0,0.82}

\begin{document}
\title{Empirical Analysis on Comparing the Performance of Alpha Miner Algorithm in SQL Query Language and NoSQL Column-Oriented Databases Using Apache Phoenix}

\author{Kunal Gupta \quad Astha Sachdev \quad Ashish Sureka\\
IIIT Delhi (IIITD)\\
       New Delhi, India\\
ABB Corporate Research\\
       Bangalore, India
}
\maketitle

\begin{abstract}
\small
Process-Aware Information Systems (PAIS) is an IT system that support business processes and generate large amounts of event logs from the execution of business processes. An event log is represented as a tuple of CaseID, Timestamp, Activity and Actor. Process Mining is a new and emerging field that aims at analyzing the event logs to discover, enhance and improve business processes and check conformance between run time and design time business processes. The large volume of event logs generated are stored in the databases. Relational databases perform well for a certain class of applications. However, there are a certain class of applications for which relational databases are not able to scale. To handle such class of applications, NoSQL database systems emerged. Discovering a process model (workflow model) from event logs is one of the most challenging and important Process Mining task. The $\alpha$-miner algorithm is one of the first and most widely used Process Discovery technique. Our objective is to investigate which of the databases (Relational or NoSQL) performs better for a Process Discovery application under Process Mining. We implement the $\alpha$-miner algorithm on relational (row-oriented) and NoSQL (column-oriented) databases in database query languages so that our algorithm is tightly coupled to the database. We present a performance benchmarking and comparison of the $\alpha$-miner algorithm on row-oriented database and NoSQL column-oriented database so that we can compare which database can efficiently store massive event logs and analyze it in seconds to discover a process model. 
\end{abstract} 


~


\section{\textbf {Research Motivation and Aim}}
\par{A PAIS is an IT system that manages and supports business processes. A PAIS generates data from the execution of business processes. The data generated by a PAIS like Enterprise Resource Planing (ERP) and Customer Relationship Management (CRM) \cite{pais} is in the form of event logs (represented as a tuple <CaseID, Timestamp, Activity, Actor>). In an event log, a particular CaseID, that is a process instance, has a set of activities associated with it, ordered by timestamp. Process Mining is new and emerging field which consist of analyzing event logs generated from the execution of business process. The insights obtained from event logs helps the organizations to improve their business processes. There are three major techniques within Process Mining \textit{viz.} Process Discovery, Process Conformance and Process Enhancement \cite{Process}. The classification of Process Mining techniques is based on whether there is a priori model and how the a priori model is used, if present. In this paper we focus on Process Discovery aspect of Process Mining. In Process Discovery, there is no a priori model. Process Discovery aims to construct a process model, which is a computationally intensive task, from the the information present in event logs. One of the most fundamental algorithm under Process Discovery is the $\alpha$-miner algorithm \cite{Sawitree} which is used to generate process model from event logs. } 
\par{Before the year 2000, majority of the organizations used traditional Relational Database Management System (RDBMS) to store the data. Most of the traditional relational databases focus on Online Transaction Processing (OLTP) applications \cite{LSuresh18} but are not able to perform certain Online Analytical Processing (OLAP) applications efficiently. Row-oriented databases are not able to perform certain analytical functions (like Dense\textunderscore Rank, Sum, Count, Rank, Top, First, Last and Average) efficiently but work fine when we need to retrieve the entire row or to insert a new record. Recent years have seen the introduction of a number of NoSQL column-oriented database systems \cite{nosql}. These database systems have been shown to perform more than an order of magnitude better than the traditional relational database systems on analytical workloads \cite{Comparison}. NoSQL column-oriented databases are well suited for analytical queries but result in poor performance for insertion of individual records or retrieving all the fields of a row. Another problem with traditional relational databases is impedance matching \cite{im}. When representation of data in memory and that in database is different, then it is known as impedance matching. This is because in-memory data structures use lists, dictionaries, nested lists while traditional databases store data only in the form of tables and rows. Thus, we need to translate data objects present in the memory to tables and rows and vice-versa. Performing the translation is complex and costly. NoSQL databases on the other hand are schema-less. Records can be inserted at run time without defining any rigid schema. Hence, NoSQL databases do not face the problem of impedance matching.} 
\par{There are certain class of applications for which row-oriented databases are not able to scale like real time messaging System of Facebook. To handle such class of applications, NoSQL database systems were introduced. Process Discovery is a very important application of Process Mining. Our aim is to examine an approach to implement a Process Discovery $\alpha$-miner algorithm on a row-oriented database and a NoSQL column-oriented database and to benchmark the performance of the algorithm on both the row-oriented and column-oriented databases. A lot of research has been done in implementing data mining algorithms in database query languages. Previous work suggests that tight coupling of the data mining algorithms to the database systems improves the performance of the algorithms significantly \cite{ron}. We aim to implement $\alpha$-miner algorithm in Structured Query Language (SQL) so that our Process Discovery application is tightly coupled to the database.} 
\par{Combination of Hadoop\footnote{http://hadoop.apache.org} component and NoSQL column-oriented databases allow accessing large data efficiently and storing data easily as compared to single machine databases \cite{LSuresh12}. There are various NoSQL column-oriented databases \cite{nosql} but for our current work, we will focus on Apache HBase\footnote{www.hbase.apache.org} (NoSQL column-oriented database) and MySQL\footnote{http://www.mysql.com/} (row-oriented database). To perform analytical functions, NoSQL column-oriented databases either use MapReduce programming model or use their own simple query language that just supports create, read, update and delete (CRUD). They do not support SQL interface. We integrate Apache Phoenix\footnote{http://www.phoenix.apache.org} (SQL layer over HBase) into HBase to support SQL interface in it. It converts SQL queries to HBase scans rather than MapReduce jobs. It executes converted scans in parallel over the regions in a regionserver and targets low latency query over HBase tables as compared to MapReduce framework and client API's.}

\par{Main research aim presented in this paper is-
\begin{enumerate}
\item To investigate an approach to implement $\alpha$-miner algorithm in SQL. The underlying row-oriented database for implementation is MySQL using InnoDB\footnote{http://dev.mysql.com/doc/refman/5.5/en/innodb-storage-engine.html} engine.
\item To investigate an approach to implement $\alpha$-miner algorithm on column-oriented database HBase using Phoenix and HDFS.
\item To conduct a series of experiment on publicly available real world dataset, to compare the performance of $\alpha$-miner algorithm on both the databases. The experiment considers multiple aspects such as $\alpha$-miner stepwise execution, bulk loading across various datasets, write intensive time, read intensive time, disk space of tables, disk space of tables using compression technique, $\alpha$-miner stepwise execution using compression technique, real time batch wise insertion and real time single record insertion.
\end{enumerate}}

\section{\textbf {Related Work}}

In this Section, we review closely related work to the study presented in this paper and list the novel contributions of our work in context to existing work. We divide related work into following three lines of research:

\normalfont \subsection {\normalfont \textbf{ Implementation of Mining Algorithms in Row-Oriented Databases}}

\par{Ordonez et al. investigate an approach to efficiently implement the EM algorithm in SQL\cite{OrdonezC}. They perform clustering of  large datasets. They effectively handle high dimensional data, a high number of clusters and more importantly, a very large number of data records. Xuequn Shang et al. present an efficient implemenation of frequent pattern mining in relational databases \cite{ron1996power1}. They propose a concept called Projection Pattern Discovery (Propad). Propad fundamentally differs from Apriori like candidate set generation-and-test approach. This approach successively projects the transaction table into frequent itemsets to avoid making multiple passes over the large original transaction table and generating a huge set of candidates.}

\normalfont \subsection {\normalfont \textbf{ Implementation of Mining Algorithms in Graph Databases, Parallelization and Utility Based Approach}}

Joishi et al. implement Similar-Task algorithm on relational and NoSQL (graph oriented) databases using only query language constructs \cite{joishiideas2015}. They conduct empirical analysis on a large real world data set to compare the performance of row-oriented database and NoSQL graph-oriented database \cite{joishiideas2015}. Kundra et al. investigate the application of parallelization on Alpha Miner algorithm \cite{kundrabpi2015}. They use Graphics Processor Unit (GPU) to run computationally intensive parts of Alpha Miner algorithm in parallel \cite{kundrabpi2015}. Anand et al. propose a Utility-Based Fuzzy Miner UBFM algorithm to efficiently mine a process model driven by a utility threshold \cite{anandbda2015}.

\subsection {\normalfont \textbf{Implementation of Mining Algorithms in Column-Oriented Databases}}

\par{Mehta et al. conducted a study of data mining algorithms on column-oriented database systems \cite{Warrender}. They study the architecture of  open source column-oriented databases and implemented tree based classification algorithm on various column-oriented databases like MonetDB and Infobright. Suresh L et al. presented an implementation of k-means clustering algorithm on column-oriented databases \cite{LSuresh}. They introduce an algorithm known as Novel Seeding Algorithm to implement k-means in column-oriented databases. This algorithm identifies the median gaps in the data in each of the columns and using these gaps to identify other clusters by using the difference in the median gaps.}

\subsection { \normalfont\textbf{Performance Comparison of Mining Algorithms in Row-Oriented and Column-Oriented Databases}}

\par{Hasso conducted common database approach for OLTP and OLAP using an in-memory column database \cite{LSuresh18}. He presented a comparison of OLAP  and OLTP considering row-oriented database and column-oriented database. Rana et al. implement Apriori algorithm on MonetDB and Oracle database and compare their performance in terms of execution time \cite{Hofmeyr}.}


\subsection{Novel Contribution}

In context of existing work, this study presented here makes the following novel contributions. The work presented in this paper is extension of the work presented in \cite{sachdevbda2015}\cite{guptac3s2e2015}. The study presented in this paper has several more results which are not present in \cite{sachdevbda2015}\cite{guptac3s2e2015} due to limited space in the conference paper.
\begin{enumerate}
\item While there has been work done on implementing data mining algorithms on row-oriented databases, we are the first to implement Process Mining $\alpha$-miner algorithm on MySQL using InnoDB storage engine.
\item While there has been work done on implementing data mining algorithms on column-oriented databases, we are the first to implement Process Mining $\alpha$-miner algorithm in HBase using Phoenix and HDFS. 
\item We present a performance benchmarking and comparison of $\alpha$-miner algorithm on both MySQL and HBase. We consider multiple aspects such as $\alpha$-miner stepwise execution, bulk loading across various datasets, write intensive time, read intensive time, disk space of tables, disk space of tables using compression technique, $\alpha$-miner stepwise execution using compression technique, real time batch wise insertion and real time single record insertion.
\end{enumerate}


\section {\textbf{$\alpha$-Miner Algorithm}}
\par{The $\alpha$-miner algorithm is an algorithm used in discovering Process Mining \cite{Sawitree}. It was first put forward by van der Aalst, Weijter and Maruster. Input for the $\alpha$-miner algorithm is an event log L and output is a process model. The $\alpha$-miner algorithm consists of scanning the event logs for discovering causality between the activities present in the event log. The basic ordering relations determined by $\alpha$-miner algorithm are the following:}
\begin{enumerate}
\item a $\succ$$_{L}$b iff a directly precedes b in some trace. Where a and b can be set of activities or an activity and this relation represents causality.

\item a $\rightarrow$$_{L}$b iff a$\succ$$_{L}$b $\wedge$ b\st{$\succ$}$_{L}$a.

\item a$\parallel$b iff a$\succ$$_{L}$b and b$\succ$$_{L}$a in some trace. 

\item a$\sharp$b iff a\st{$\succ$}$_{L}$b $\wedge$  b\st{$\succ$}$_{L}$a.

\end{enumerate}
\par{Let L be an event log over T, where T is the set of distinct activities present in the event log and $\sigma$ is a trace in the event log. $\alpha$(L) is defined as follows.}








The stepwise description of the $\alpha$-miner algorithm can be given as:
\begin{enumerate}
\item Step $1$ computes T$_{L}$ (Total Events) which represents the set of distinct activities present in the event log L.
\item Step $2$ computes T$_{I}$ (Initial Events) which represents the set of all the initial activities of corresponding trace.
\item Step $3$ computes T$_{O}$ (Final Events) which represents the set of distinct activities which appear at the end of some trace in the event log.
\item In order to compute Step $4$, we compute the relationships between all the activities in T$_{L}$. This computation is presented in the form of a footprint matrix and is called pre-processing in $\alpha$-miner algorithm. Using the footprint matrix we compute pairs of sets of activities such that all activities in the same set are not connected to each other while every activity in first set has causality relationship to every other activity in the second set.
\item Step $5$ keeps only the maximal pairs of sets generated in the fourth step, eliminating the non-maximal ones. 
\item Step $6$ adds the input place which is the source place and the output place which is the sink place in addition to all the places obtained in the fifth step.
\item Step $7$ is the final step of the $\alpha$-miner algorithm that presents all the places including the input and output places and all the input and output transitions from the places. 
\end{enumerate}

\section{Implementation of $\alpha$-Miner Algorithm in SQL on Row-Oriented Database (MySQL)}
We present a few segments of our implementation due to limited space in the paper. The entire code and implementation can be downloaded from our website\footnote{https://dl.dropboxusercontent.com/u/48972351/Programing-Alpha-miner-in-MySQL-and-HBase.zip}. Before implementing $\alpha$-miner algorithm, we do pre-processing in JAVA to create two tables \textit{viz.} causality table (consist of two column eventA and eventB) and notconnected table (consist of two column eventA and eventB).
\begin{enumerate}

\item We create a table eventlog using create table\footnote{http://dev.mysql.com/doc/refman/5.1/en/create-table.html} keyword consisting of 5 columns (CaseID, Timestamp, Status, Activity and Actor) each of which are varchar datatype except Timestamp which is of timestamp datatype. The primary key is a composite primary key consisting of CaseID, Timestamp and Status.

\item  We load the data into table eventlog using LOAD DATA INFILE\footnote{http://dev.mysql.com/doc/refman/5.1/en/load-data.html} command.

\item  For Step 1, we create a table totalEvent that contains a single column (event) which is of varchar datatype. To populate the table we select distinct activities from the table eventlog.

\item  For Step 2, we create a table initialEvent that contains a single column (initial) which is of varchar datatype. To populate the table
     \begin{enumerate}
     \item We first select the minimum value of Timestamp from table eventlog by grouping CaseID.
     \item Then we select distinct activities from table eventlog for every distinct value of CaseID where Timestamp is the minimum Timestamp.
     \end{enumerate}
     
\item  For Step 3, we create a table finalEvent that contains a single column (final) which is of varchar datatype. To populate the table
     \begin{enumerate}
     \item We first select maximum Timestamp from a table eventlog by grouping CaseID.
     \item Then we select distinct activities from a table eventlog for every distinct value of CaseID where Timestamp is the maximum Timestamp. 
     \end{enumerate}  
     
\item For Step 4, we create five tables \textit{viz.} SafeEventA, SafeEventB, EventA, EventB and XL. All the five tables contain two columns (setA and setB) which are of varchar datatype.  
\begin{enumerate}
     \item In table causality we use group\textunderscore concat\footnote{http://dev.mysql.com/doc/refman/5.0/en/group-by-functions.html\#function\textunderscore group-concat} to combine the values of column eventB of corresponding value of a column eventA and insert the results in a table EventA. 
     \item In table causality we use group\textunderscore concat to combine the values of column eventA of corresponding value of a column eventB and insert the results in the table EventB. 
     \item To populate tables SafeEventA and SafeEventB-
      \begin{enumerate}
      \item Select setA and setB from tables EventA and EventB
      \item For every value of setB in table EventA, if value is present in table notconnected, insert the corresponding value of setA and setB in table SafeEventA. Repeat the same step for populating table SafeEventB.
      \end{enumerate}
     \item To populate table XL, we insert all the rows from the three tables SafeEventA, SafeEventB and causality.
     \end{enumerate}  
     
\item For Step 5, we create three tables \textit{viz.} eventASafe, eventBSafe and YL. All the three tables contain two columns (setA and setB) which are of varchar datatype.    
\begin{enumerate}
\item We create a stored procedure to split the values of column setB of table SafeEventA on comma separator. Insert the results in safeA table. 
\item We create a stored procedure to split the values of column setA of table SafeEventB on comma separator. Insert the results in safeB table. 
\item To populate table eventASafe, insert all the rows from table safeA. 
\item To populate table eventBSafe, insert all the rows from table safeB. 
\item To populate table YL, insert all the rows from tables SafeEventA, SafeEventB, eventASafe, eventBSafe and causality.
\end{enumerate}
\item For Step 6, we create two tables \textit{viz.} terminalPlace that contains a single column (event) which is of varchar datatype and PL which also contains a single column (Place) which is of varchar datatype. 
     \begin{enumerate}
     \item To populate table terminalPlace, insert 'i' and 'o' in the table.
     \item To populate table PL, we use concat\textunderscore ws \footnote{http://dev.mysql.com/doc/refman/5.0/en/string-functions.html\#function\textunderscore concat-ws} to combine the values of column setA and column setB of a table YL using \& separator and insert the results in table PL. Furthermore, we insert all the rows of table terminalPlace into table PL.
     \end{enumerate}

\item For Step 7, we create 3 tables \textit{viz.} Place1 and Place2 which consist of two columns (id and value) which are of varchar datatype and FL which consists of two columns (firstplace and secondplace) which are of varchar datatype.
\begin{enumerate}
\item To populate table Place1, we use concat\textunderscore ws to combine the values of column setA and column setB of table YL using \& separator. Insert the results in column setB of table Place1. Insert all the values of column setA of table YL into column setA of table Place1. 
\item To populate table Place2, we use concat\textunderscore ws to combine the values of column setA and column setB of  table YL using \& separator. Insert the results in column setA of table Place2. Insert all the values of column setB of table YL in column setB of table Place2. 
\item We create a stored procedure to split column setB of table Place1 on comma separator. In stored procedure we create  table temp\textunderscore place2 to insert the results.
\item We create a stored procedure to split column setA of a table Place2 on comma separator. In stored procedure we create table temp\textunderscore place2 to insert the results.
\item To populate a table FL, insert all the rows from tables temp\textunderscore place1 and temp\textunderscore place2. Insert the results of cross join of two tables \textit{viz.} terminalPlace and intialEvent and of table finalEvent and table terminalPlace.
\end{enumerate}
\end{enumerate}

\section{Implementation of $\alpha$-Miner Algorithm on NoSQL Column-Oriented Database (HBase) Using Apache Phoenix}
Before implementing $\alpha$-miner algorithm, we do pre-processing in JAVA to create two tables \textit{viz.} causality table (consist of two column eventA and eventB) and notconnected table (consist of two column eventA and eventB).

\begin{enumerate}
\item To create table eventlog (Refer section 4 point 1). To load the data in table eventlog, we use MapReduce framework\footnote{http://phoenix.apache.org/bulk\textunderscore dataload.html}. 
\item For Step 1, 2 and 3 (Refer section 4 point 3, point 4 and point 5)
\item For Step 4, we create three tables \textit{viz.} SafeEventA, SafeEventB and XL. All the three tables consist of two columns (setA and setB) which are of varchar datatype.  
\begin{enumerate}
     \item {\fontsize{10}{10}\selectfont
     \begin{algorithm}[H]
\SetAlgoLined
Select eventA, eventB from causality.\\
Select setA, setB from notconnected.\\
Compare Function that compares whether set of activities is notconnected.\\
\ForEach { \textnormal{eventA}  \textnormal {in the causality}}{
    Form single group say \emph{grp} of all activity present in column \textbf{eventB}. Pass \emph{grp} to \textbf{Compare} function. For any such combination returning true, insert \textbf{eventA} in\textbf{ setA }and that combination into \textbf{setB} of table \textbf{SafeEventA}.
}
\ForEach { \textnormal{eventB}  \textnormal {in the causality}}{
   Form single group say \emph{grp} of all activity present in column \textbf{eventA}. Pass \emph{grp} to \textbf{Compare} function. For any such combination returning true, insert that combination into \textbf{setA} and \textbf{eventB} in \textbf{setB} of table \textbf{SafeEventA}.
}
 \caption{Populating table SafeEventA and SafeEventB }
\end{algorithm}
}
     \item To populate table XL, we insert all the rows from three tables SafeEventA, SafeEventB and causality.
     \end{enumerate}  
     
\item For Step 5, we create three tables \textit{viz.} EventA, EventB and YL. All the three tables consist of two columns (setA and setB) which are of varchar datatype.    

\begin{enumerate}
     \item {\fontsize{10}{10}\selectfont
     \begin{algorithm}[H]
\SetAlgoLined
Select setA, setB from SafeEventA.\\
Select setA, setB from SafeEventB.\\
\ForEach { \textnormal{setA,setB}  \textnormal {in the SafeEventA}}{
    Delimit value of setB. For all such value setB$_{i}$, insert setA and setB$_{i}$ in table \textbf{EventA}.
}
\ForEach { \textnormal{setA,setB}  \textnormal {in the SafeEventB}}{
   Delimit value of setA. For all such value setA$_{i}$, insert setA$_{i}$ and setB in table \textbf{EventB}.
}
 \caption{Populate table EventA and EventB}
\end{algorithm}
}
     \item To populate table YL, we insert all the rows from three tables EventA, EventB and causality.
     \end{enumerate}
     
\item For Step 6 (Refer section 4 point 8).
     
\item For Step 7, we create table FL that consists of two columns (Place1 and Place2) which are of varchar datatype.
\begin{enumerate}
\item {\fontsize{10}{10}\selectfont
\begin{algorithm}[H]
\SetAlgoLined
Select setA, setB from YL.\\
Select final from FinalEvents.\\
Select initial from InitialEvents.\\
\ForEach { \textnormal{final}  \textnormal {in the FinalEvents}}{
 Insert final in column \textbf{Place1} and 'o' in column \textbf{Place2 }of table \textbf{FL}
}
\ForEach { \textnormal{initial}  \textnormal {in the InitialEvents}}{
 Insert 'i' in column \textbf{Place1} and \textbf{initial} in column \textbf{Place2} of table \textbf{FL}
}
\ForEach { \textnormal{setA,setB}  \textnormal {in the YL}}{
 If value of \textbf{column setA} has set of activities instead of single activity then delimit. \textbf{Each split value will be stored in column Place1} and \textbf{combination of setA and setB in column Place2} of table \textbf{FL}\\
 else choose column \textbf{setB} and delimit. \textbf{Each split value will be stored in column Place2} and \textbf{combination of setA and setB in column Place1} of table \textbf{FL}
}
 \caption{Populating Table \textbf{FL}}
\end{algorithm}
}
\end{enumerate}
\end{enumerate}

\section{\textbf{Experimental Dataset}}

\par{We conduct our study on a publicly available large real world dataset downloaded from Business Process Intelligence $2014$ (BPI $2014$)\footnote{http://www.win.tue.nl/bpi/2014/start}. The dataset is provided by Robobank Information and Communication and Technology (ICT). The data is related to Information Technology Infrastructure Library (ITIL) process implemented in the bank. ITIL is a process which starts when a client reports an issue regarding disruption of ICT service to Service Desk Agent (SDA). SDA records the complete information about the problem in an Interaction record. If the issue does not get resolved on first contact then an Incident record is created for the corresponding Interaction else the issue is closed. If an issue appears frequently then a request for change is initiated. Robobank provides $4$ files in CSV format \textit{viz.} Change records, Incident records, Interaction records and Incident activity records. We import Incident activity records CSV file in MySQL and HBase for benchmarking and performance comparison of $\alpha$-miner algorithm. Incident activity records file contains $4,66,738$ number of records and contains the following fields \textit{viz.} Incident ID, DateTimeStamp, IncidentActivity\textunderscore number, IncidentActivity\textunderscore Type, Interatcion ID, Assignment Group and KM Number. Out of these we use the following fields: 

\begin{enumerate}
\item Incident ID: The unique ID of a record in the Service Management tool. It is represented as CaseID in our data model.
\item DateTimeStamp: Date and time when a specific activity starts. It is represented as timestamp in our data model.
\item IncidentActivity\textunderscore Type: Identifies which type of an activity takes place.
\item Assignment Group: The team responsible for an activity.
\end{enumerate}


\section{\textbf{Benchmarking and Performance Comparison}}

\par{Our benchmarking system consists of Intel Core i$3$ $2.20$ GHz processor, $4$ GB Random Access Memory (RAM), $500$ GB Hard Disk Drive (HDD), Operating System (OS) is Linux Ubuntu $14.04$ LTS and Cache of $3$ MB. The experiments were conducted on MySQL $5.6$ (row-oriented database) and HBase $0.96.1$ (NoSQL column-oriented database) with HDFS $2.3.0$ as the file system below it and a layer of Phoenix $4.2.1$ above it. We conduct series of experiments on a single machine.} 
\par{The $\alpha$-miner algorithm interacts with the database. The underlying data model for implementing $\alpha$-miner algorithm consists of $5$ columns (CaseID, Timestamp, Status, Activity and Actor) each of which are varchar datatype except Timestamp which is of timestamp datatype. The primary key is a composite primary key consisting of CaseID, Timestamp and Status. We use the same data model while performing bulk loading of datasets through the database loader. We take each reading five times for all the experiments and the average of each reading is reported in the paper.}

\begin{table}[H]
\captionsetup{font=scriptsize}
\CenterFloatBoxes
\begin{floatrow}
\ttabbox
  { \begin{tabular}{|c|c|c|}
  \hline
  
      \textbf{Dataset Size} & \multicolumn{2}{|c|}{\textbf{Load Time in Seconds}}\\
    \hline
    \centering
    & \textbf{MySQL}& \textbf{HBase}\\ \hline
    
     1,00,000  & 12.98  &  12.03    \\ \hline
     4,00,000  & 46.79  &  42.94    \\ \hline
     8,00,000  & 156.79  &  64.48    \\ \hline
     12,00,000 & 3654.14 &   89.55   \\ \hline
     16,00,000 & 8408.20  &  123.85   \\ \hline
     20,00,000 &    13536.42&       145.53 \\ \hline  
     
  \end{tabular}
  }
  {\caption{Dataset Load Time}}
 \ttabbox
  { \begin{tabular}{|c|c|c|}
  \hline

       \textbf{Steps} & \multicolumn{2}{|c|}{ \textbf{Execution Time in Seconds}}\\
    \hline
       \centering
    & \textbf{MySQL}& \textbf{HBase}\\ \hline
     1 & 4.19 & 2.89 \\ \hline
2 & 6.29 & 5.82 \\ \hline
3 & 6.71 & 5.74 \\ \hline
4 & 4.09 & 3.89 \\ \hline
5 & 8.23 & 5.64 \\ \hline
6 & 2.04 & 1.00 \\ \hline
7 & 9.18 & 3.09  \\ \hline

  \end{tabular}
  }
 {\caption{Stepwise Execution Time}}
\end{floatrow}
\end{table}


\begin{figure}[h]
  \centering
  \mbox{
    \subfigure[Dataset Load Time in Seconds]{\includegraphics[width=8cm,height=4cm]{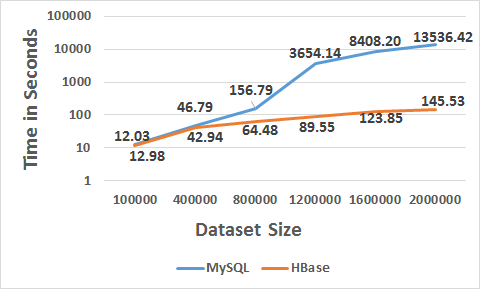}}\quad
    \subfigure[$\alpha$-miner Stepwise Execution Time in Seconds]{\includegraphics[width=7cm,height=4cm]{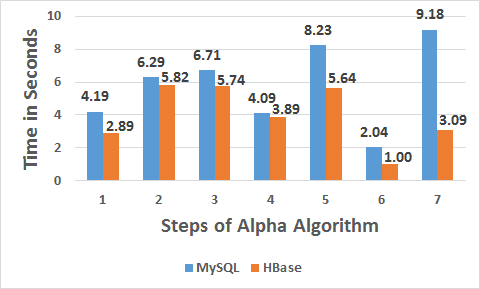}} \quad
   
  }
  \caption{Dataset Load Time and $\alpha$-miner Stepwise Execution Time }
\end{figure}

\par{Our first experiment consists of investigating the time taken to perform bulk loading in both the databases across various dataset sizes. Table 1 shows that the average time taken to load data in HBase is $29$ times lower as compared to MySQL. Bulk loading in HBase is done using the MapReduce framework. Phoenix has an inbuilt script of MapReduce using which we conduct our experiment. We use two mappers and two reducers for running MapReduce jobs. The script requires two parameters before running MapReduce \textit{viz.} input file and output file. Input file must be present in HDFS and script creates empty output file in HDFS after executing it. Due to parallelism, all the key-values of the input file are mapped to two mappers and the output of each mapper is passed to two reducers. MapReduce converts all the data of the input file into the format of HFiles (HBase file format) before it handovers to HBase. HFile stores data in key-value pairs and reducers also generate output in key-value pairs. The output of reducers can be stored on multiple HFiles directly without interacting with HBase. At the end all the created HFiles will be handovered to HBase to store on HDFS. Bulk loading in MySQL is done using LOAD DATA INFILE command which is designed for mass loading of records in a single operation as it overcomes the overhead of parsing and flushing batch of inserts stored in a buffer to MySQL server. LOAD DATA INFILE command also creates an index and checks uniqueness while inserting records in a table. Therefore, in case of MySQL, while inserting large datasets, most of the time is spent in checking uniqueness and creating indexes. Fig. 1(a) reveal that when the dataset size increases then the difference between the time taken in loading data in MySQL and HBase also increases. The performance of HBase is better as compared to MySQL because the percentage increase of time in MySQL is $3.5$ times more as compared to HBase.}

\par{$\alpha$-miner algorithm is a seven step algorithm (Refer to Section $3.2$). Few steps are read intensive (Steps 1,2,3) while few steps are write intensive (Steps 4,5,6,7). We perform an experiment to compute the $\alpha$-miner algorithm execution time of each step in both MySQL and HBase to examine which database performs better for each step. In MySQL default size of innodb\textunderscore buffer\textunderscore pool\textunderscore size is 8 MB that is used to Cache data and indexes of its tables. The larger we set this value, the lesser is the disk I/O needed to access the data in tables. Table 2 and Fig. 1(b) reveal that the the stepwise time taken in HBase is always lower as compared to MySQL for all the Steps. We conjecture the reason for HBase performing better than MySQL can be the difference in the internal architecture of MYSQL and HBase. For the first three steps, both MySQL and HBase perform full table scans. In case of MySQL, the entire row is first retrieved sequentially and then the specific attributes are retrieved. However, in case of HBase, table is stored on multiple regions and Phoenix performs parallelism on multiple regions of a table leading to better performance of HBase in comparison to MySQL. Furthermore, in HBase, only the specific attributes specified in the query are retrieved. The overhead of retrieving the entire row is not present in HBase. Hence, HBase gives a better performance for the first three steps.}
\par{The remaining steps read data from the tables obtained in the first three steps and write it to the tables created during their execution. In MySQL, in order to read the data from a table we need to scan the B-Tree index to find the location of block where data is stored. In case of HBase data is read from the memstore. If values are not in memstore they are read from HDFS. Thus, the read performance of HBase is better as compared to MySQL. Similarly, in MySQL, in order to write data, the entire B-Tree index needs to be scanned to locate the block where we need to write data. HBase follows log structure merge tree index. In case of HBase, values are written in append only mode. The writes in HBase are sequential because first it writes to WAL (Write Ahead Log) of regionserver and then to memstore of corresponding region. HBase lags in persisting data to disk. Hence, HBase gives better write performance as compared to MySQL. Therefore, the total execution time of $\alpha$-miner algorithm in HBase is $1.44$ times lower than that of MySQL.}


\begin{table}[H]
\captionsetup{font=scriptsize}
\CenterFloatBoxes
\begin{floatrow}
\ttabbox
  { \begin{tabular}{|c|c|c|}
  \hline
   
      \textbf{Steps} & \multicolumn{2}{|c|}{\textbf{Read Time in Seconds}}\\
    \hline
    & \textbf{MySQL}& \textbf{HBase}\\ \hline
    1 &  2.06 &  1.60    \\ \hline  
    2 &  4.78 &  4.48  \\ \hline
     3 &  4.95 &  4.70    \\ \hline
     4 &  1.36 &  1.29    \\ \hline
     5 &  0.37 &  0.37    \\ \hline
     6 &  0.12 &  0.11   \\ \hline
     7 &  1.06 &  0.19    \\ \hline
  \end{tabular}
  }
  {\caption{Read Intensive Time}}
 \ttabbox
  { \begin{tabular}{|c|c|c|}
  \hline
       \textbf{Steps} & \multicolumn{2}{|c|}{ \textbf{Write Time in Seconds}}\\
    \hline
    &\textbf{MySQL}&\textbf{HBase}\\ \hline
     1  &  2.12& 1.29   \\ \hline  
      2 &  1.50& 1.33  \\ \hline
      3 &  1.76& 1.04    \\ \hline
      4 &  2.73& 2.59  \\ \hline
      5 &  7.85& 5.27   \\ \hline
      6 &  1.92& 0.89  \\ \hline
      7 & 8.12 & 2.90   \\ \hline
      
  \end{tabular}
  }
 {\caption{Write Intensive Time}}
\end{floatrow}
\end{table}
  \begin{figure}[h]
  \centering
  \mbox{
    \subfigure[Read Intensive Time in Seconds]{\includegraphics[width=8cm,height=4cm]{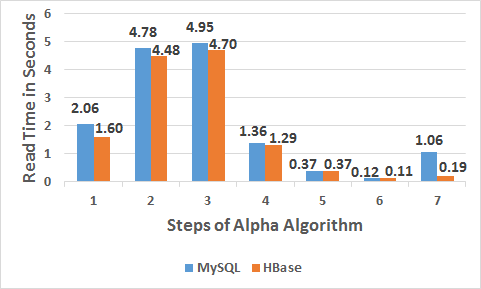}}\quad
    \subfigure[Write Intensive Time in Seconds]{\includegraphics[width=7cm,height=4cm]{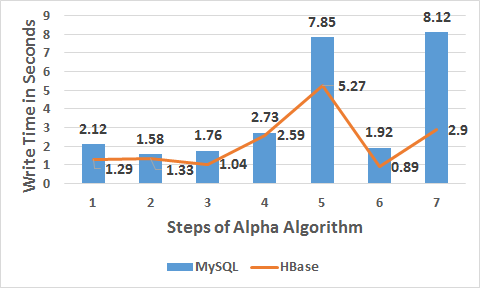}} \quad
   
  }
  \caption{Read and Write Intensive Time }
\end{figure}


\par{$\alpha$-miner algorithm consists of seven steps (Refer to Section $3.2$). Few steps are more intensive for read operations while few steps are more intensive for write operations. We conduct an experiment to compare which of the database performs better for read and write operations in both MySQL and HBase. As can be seen from Fig. 2(a) and Table 3, HBase gives better read performance as compared to MySQL for all the Steps. According to us, the reason for HBase giving better read performance can be the difference in the data structure of both the databases. In MySQL, B-Tree index needs to be scanned to find the location of block where the data is stored. In case of HBase data is read as described below- 
\begin{enumerate}
\item To find the data, HBase client will hit the memstore first.
\item When the memstore fails, HBase client will hit the BlockCache \cite{Hbase}.
\item If both the memtsore and BlockCache fail, HBase client will locate the target HFiles in HDFS (contains target data) using log structure merge tree and load it into the memory.
\end{enumerate}
}
\par{The total time taken to read the data in each of the Step of $\alpha$-miner algorithm is $1.16$ times lower in HBase as compared to MySQL. Fig. 2(b) and Table 4 show that the write performance of HBase is better as compared to MySQL. We believe the reason for HBase giving better write performance can be the difference in the way writes are performed in both the databases. In MySQL, the B-Tree index needs to be scanned to find the location of block where the data needs to be written. Almost all the leaf blocks of B-Tree are stored on the disk. Hence, at least one I/O operation is required to retrieve the target block in memory. Fig. 2(b) illustrates that Step 5 and Step 7 of $\alpha$-miner algorithm in MySQL are more write intensive than the other steps. We believe the reason can be the generation of maximal sets and places by stored procedures in MySQL. A large number of insert operations are executed in the stored procedure to generate the maximal sets. In HBase we perform the same steps using Java because SQL interface over HBase does not support advanced features of SQL. Writes in HBase are performed by first locating regionserver from zookeeper\footnote{http://www.zookeeper.apache.org}, then regionserver writes to WAL and finally to memstore of the corresponding region. Phoenix allows to perform parallelism in reading and writing the data on multiple regions of a table stored in HBase regionserver in comparison to sequential reads and writes of MySQL. The total time taken in writing the data in each of the Step of $\alpha$-miner algorithm is $1.70$ times lower in HBase as compared to MySQL. Thus, writes in HBase are more optimized as compared to that in MySQL.}

  
  \begin{table}[H]
  \captionsetup{font=scriptsize}
\CenterFloatBoxes
\begin{floatrow}
\ttabbox
  {\begin{tabular}{|c|c|c|}
  \hline
       \textbf{Step wise Tables} & \multicolumn{2}{|c|}{ \textbf{Disk Usage in Bytes}}\\
    \hline
    &\textbf{MySQL}&\textbf{HBase}\\ \hline
    Step 1 & 16384 & 2048 \\ \hline
    Step 2 & 16384 & 1945 \\ \hline
    Step 3 & 16384 & 1945 \\ \hline
    Step 4 & 16384 & 6348 \\ \hline
    Step 5 & 16384 & 3481 \\ \hline
    Step 6 & 16384 & 4505\\ \hline
    Step 7 & 49152 & 13414\\ \hline
      
  \end{tabular}
  }
 {\caption{Disk Usage of Tables}}
 \ttabbox
  { \begin{tabular}{|c|c|c|}
  \hline
       \textbf{Step wise Tables} & \multicolumn{2}{|c|}{ \textbf{Disk Usage in Bytes}}\\
    \hline
    &\textbf{MySQL}&\textbf{HBase}\\ \hline
Step 1 & 8192 & 1536 \\ \hline
Step 2 & 8192 & 1433 \\ \hline
Step 3 & 8192 & 1433 \\ \hline
Step 4 & 8192 & 2355 \\ \hline
Step 5 & 8192 & 1843 \\ \hline
Step 6 & 8192 & 1945\\ \hline
Step 7 & 8192 & 3584\\ \hline
      
  \end{tabular}
  }
  {\caption{Disk Usage of Tables With Compression}}
\end{floatrow}
\end{table}
  
  
  \begin{figure}[h]
  \centering
  \mbox{
    \subfigure[Disk Usage of Tables in Bytes]{\includegraphics[width=8cm,height=4cm]{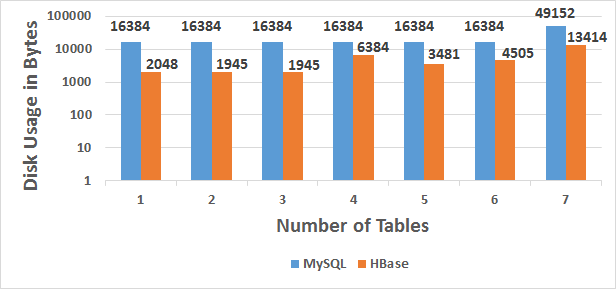}}\quad
    \subfigure[Disk Usage of Tables in Bytes with Compression]{\includegraphics[width=7cm,height=4cm]{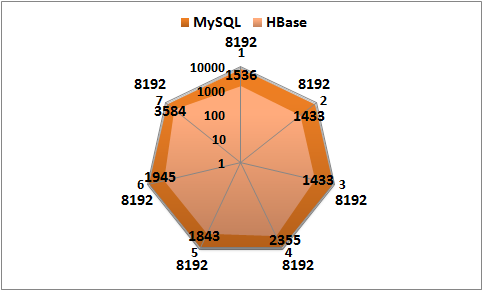}} \quad
   
  }
  \caption{Disk Usage of Tables in Bytes with and without Compression }
\end{figure}
 

\par{We perform an experiment to investigate which database can efficiently store results of each Step of $\alpha$-miner algorithm in tables with minimum disk space. Table 5 and Fig. 3(a) reveal the disk space occupied by tables created in each step of $\alpha$-miner algorithm. We include only data length (excluding the size of index tables) in disk space of table because we did not create index for any of the tables. Experimental results show that HBase on an average uses disk space $6$ times lower than MySQL for tables created at each step of the algorithm. Hence, cumulative disk space for storing all the tables in MySQL is $147456$ bytes while for HBase is $33722$ bytes. We believe the underlying reason for MySQL occupying more space is the difference in the way memory is allocated to tables in both the databases. In MySQL, the operating system allocates fixed size blocks of size $16$ KB for the data to be stored in a table. Number of blocks assigned to a table is computed by dividing the dataset size by the block size. In MySQL if set of blocks or one block has been allocated for a table then that set of blocks or block can be used only by that table. Either data in a table completely utilizes the space of all blocks or the space of the last block is unutilized. Storing smaller size file (< $16$ KB) in $16$ KB block leads to under utilization of space and the remaining space cannot be utilized by other files.}
\par{HFile is a file format of HBase which is stored over HDFS block (default size is $64$ MB). Maximum size of a HFile is 64 KB after which a new HFile needs to be created. HFiles are created when memstore reaches its threshold value (default value is $64$ MB) or commit occurs. When memstore reaches its threshold value it flushes 64 MB data of key-value pairs and creates $1024$ numbers of HFiles. If commit occurs before it reaches the threshold value then it flushes only that amount of data present in a memstore. HFile size will be equivalent to flushed amount of data from memstore. HDFS allocates blocks for incoming files by dividing the file size with the block size. For example, we have a system with $300$ MB HDFS block size. To store a $1100$ MB file, HDFS will break that file into three $300$ MB blocks and one $200$ MB block size and store it on the datanodes. The $200$ MB file is not exactly divisible by $300$. Therefore, the final block of the file is sized as modulo of the file size by block size, i.e a $200$ MB block size. Similarly, the same process is applied to the HFiles of HBase for storing in HDFS. We conclude that the disk space for each table created in each step is more efficiently utilized in HBase as compared to MySQL.}

\par{ A way to utilize disk space efficiently is by using the well known compression technique. Data compression enables smaller database size, reduced I/O and improved throughput. We conduct an experiment to compute the disk space occupied by tables at each Step of the $\alpha$-miner algorithm using compression technique. When we compare the disk space occupied by each table without compression and with compression technique we observe that the compression ratio (Actual size of table/Compressed size of table) is better in MySQL as compared to HBase. As can be seen from Table $5$ and Table $6$, the compression ratio in MySQL for Step $7$ is equal to $6$ ($49152$/$8192$) while the compression ratio in HBase for Step $7$ is equal to $3.7$. Minimum and maximum compression ratio in HBase is 1.3 and 3.7 respectively while in MySQL is 2 and 6 respectively. We believe the reason for MySQL having a higher compression ratio can be the difference in the compression techniques used by both the databases. MySQL uses the zlib compression technique which provides a better compaction using only six bytes of header and trailer of compressed block. HBase uses gzip compression technique and gzip wrapper uses a minimum of eighteen bytes of header and trailer for compressed block. The maximum compression ratio provided by MySQL is 2 times more as compared to HBase. In the context of $\alpha$-miner algorithm, MySQL performs better than HBase in utilizing the disk space when compression technique is applied.}

  
\begin{table}[H]
\captionsetup{font=scriptsize}
\CenterFloatBoxes
\begin{floatrow}
\ttabbox
  { \begin{tabular}{|c    |c|c|}
  \hline

       \textbf{Steps} & \multicolumn{2}{|c|}{ \textbf{Execution Time in Seconds}}\\
    \hline
       \centering
    & \textbf{MySQL}& \textbf{HBase}\\ \hline
     1 & 9.95 & 3.02 \\ \hline
2 & 12.96 & 6.87 \\ \hline
3 & 12.35 & 6.92 \\ \hline
4 & 5.15 & 4.12 \\ \hline
5 & 9.82 & 6.04 \\ \hline
6 & 2.62 & 2.01 \\ \hline
7 & 12.42 & 3.43 \\ \hline
  \end{tabular}
  }
 {\caption{Stepwise Execution Time with Compression}}
 \ttabbox
   { \begin{tabular}{|c    |c    |c  |}
  \hline
   
      \textbf{\specialcell[c]{ Batch Size\\ for 500 Thousand\\ Records} } & \multicolumn{2}{|c|}{\textbf{\specialcell[c]{ Batch \\wise Insertion \\Time in Seconds }}}\\
    \hline
    & \textbf{MySQL}& \textbf{HBase}\\ \hline
    30,000 &   522&  25    \\ \hline  
    60,000 &  529 &  28  \\ \hline
     90,000 &  523 &  30    \\ \hline
     1,30,000 &  527 &  32    \\ \hline
     2,00,000 &  519 &  32    \\ \hline
     2,50,000 &  527 &  32    \\ \hline
     5,00,000 &  527 &  34    \\ \hline
  \end{tabular}
  }
  {\caption{Batch wise Insertion Time}}
\end{floatrow}
\end{table}


\begin{figure}[h]
  \centering
  \mbox{
    \subfigure[$\alpha$-miner Stepwise Execution Time with Compression]{\includegraphics[width=8cm,height=4cm]{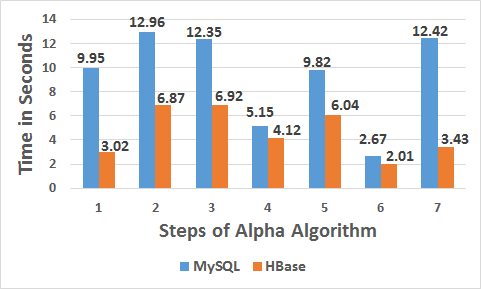}}\quad
    \subfigure[Batch wise Insertion Time in Seconds]{\includegraphics[width=7cm,height=4cm]{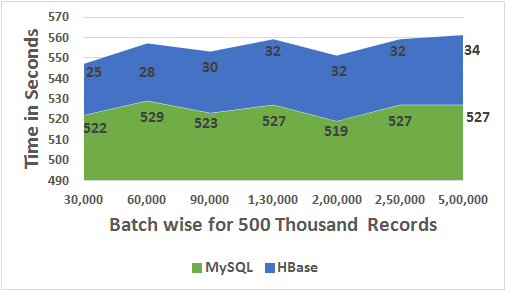}} \quad
   
  }
  \caption{$\alpha$-miner Stepwise Execution Time with Compression and Batch wise Insertion Time in Seconds  }
\end{figure}


\par{We conduct an experiment to examine the time taken by each Step of $\alpha$-miner algorithm with compression technique. In $\alpha$-miner algorithm we create tables in each Step with the compression keyword. Table 7 and Fig. 4(a) illustrate that the performance of HBase is better as compared to that of MySQL for each Step of $\alpha$-miner algorithm. We believe the reason for HBase giving better step wise execution time, with compression enabled can be the difference in the way compression is performed in both the databases. MySQL uses a block size of 1 KB, 2 KB, 4 KB, 8 KB and 16 KB. The default block size after compression in MySQL is 8 KB. Suppose the size of the compressed block is 5 KB. The block will then be uncompressed, split into two blocks and then recompressed into blocks of size 4 KB and 1 KB. All the data in a table is stored in blocks comprising a B-Tree index. The compression of B-Tree blocks is handled differently because they are frequently updated. It is important to minimize the number of times B-Tree blocks are split, uncompressed and recompressed. MySQL maintains system information in B-Tree block in uncompressed form for certain in-place updates. MySQL avoids unnecessary uncompression and recompression of blocks when they are changed because it causes latency and degrades the performance. HBase does not have fixed block size constraint after compressing the block. We conjecture that another reason for HBase giving a better stepwise execution time, with compression enabled can be the difference in the internal architecture of both the databases that was explained in experiment (Refer to Table 2 and Fig. 1(b)). From Table 2 and Table 7, we infer that the total execution time of $\alpha$-miner algorithm in MySQL is $2$ times more as compared to HBase using compression technique. We compare the total time taken in executing $\alpha$-miner algorithm without compression and with compression technique in MySQL and HBase. We observe that total time taken in executing $\alpha$-miner algorithm by HBase without compression technique is $1.33$ times lower than HBase with compression technique. Similarly, MySQL without compression technique is $1.60$ times lower than MySQL with compression technique.}
\par{In all the experiments described above the event logs generated from business processes is stored in a CSV file and then loaded in the database. In the context of Process Mining, PAIS are getting continuously updated with event logs. We setup our experiment to import the event logs directly into the database server from a client application, that is real time data (event logs) loading. The real time loading experiment can be conducted in two ways \textit{viz.} batch insertion and single row insertion. In the batch insertion, the client application inserts $5,00,000$ records in different batch sizes. The results of batch insertion are shown in Fig. 4(b) and Table 8. We believe that batch insertion might be faster than single record insertion because when we execute a batch, then multiple records in a batch are inserted in a table in a single round trip.}

\begin{table}[H]
\captionsetup{font=scriptsize}
\CenterFloatBoxes
\begin{floatrow}
\ttabbox
  { \begin{tabular}{|c|c|c|}
  \hline
       \textbf{\specialcell[c]{ Batch wise\\ for 500 Thousand\\ Records}} & \multicolumn{2}{|c|}{ \textbf{\specialcell[c]{Number of Inserts \\per Second\\ }}}\\
    \hline
    &\textbf{MySQL}&\textbf{HBase}\\ \hline
     30,000  & 957 & 19614   \\ \hline  
      60,000 & 944 & 17498   \\ \hline
      90,000 & 955 & 15340    \\ \hline
      1,30,000 & 947 & 15134  \\ \hline
      2,00,000 & 962 & 15090   \\ \hline
      2,50,000 & 948 & 15065    \\ \hline
      5,00,000 & 947 & 14613   \\ \hline
      
  \end{tabular}
  }
 {\caption{Number of Inserts per Second in Batch}}
  \ttabbox
  {\begin{tabular}{|c|c|c|}
  \hline
       \textbf{Dataset Size} & \multicolumn{2}{|c|}{ \textbf{\specialcell[c]{Single \\Row Insertion \\Time in Seconds }}}\\
    \hline
    &\textbf{MySQL}&\textbf{HBase}\\ \hline
   30,000  & 38 & 5   \\ \hline  
      60,000 & 68 & 8   \\ \hline
      90,000 & 95 & 9    \\ \hline
      1,30,000 & 134 & 10  \\ \hline
      2,00,000 & 202 & 16   \\ \hline
      2,50,000 & 255 & 18    \\ \hline
      5,00,000 & 523 & 39   \\ \hline
      
  \end{tabular}
  }
  {\caption{Single Row Insertion Time}}
\end{floatrow}
\end{table}


\begin{figure}[h]
  \centering
  \mbox{
    \subfigure[Number of Inserts per Second in Batch  ]{\includegraphics[width=8cm,height=4cm]{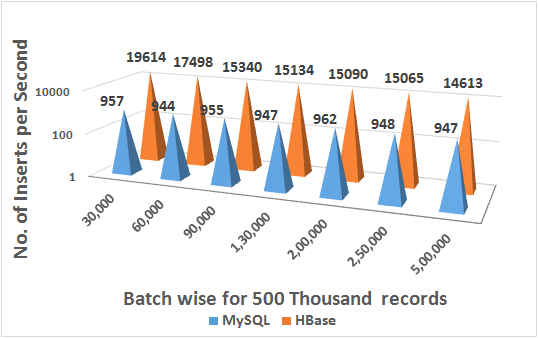}}\quad
    \subfigure[Single Row Insertion Time in Seconds ]{\includegraphics[width=7cm,height=4cm]{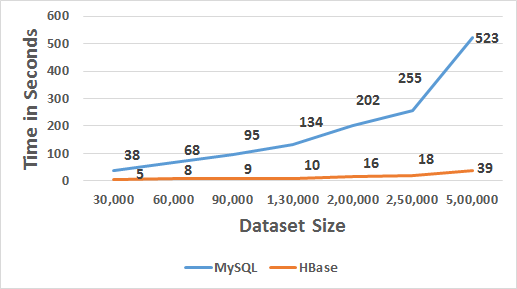}} \quad
   
  }
  \caption{Number of Inserts per Second in Batch and Single Row Insertion Time in Seconds }
\end{figure}


\par{Within the batch insertion experiment we find the number of inserts per second for different batch sizes. We calculated inserts per second by dividing total inserts with the total time taken in seconds. Fig. 5(a) illustrates that the number of inserts per second decreases as batch size increases in HBase while in MySQL it remains constant. On an average, number of inserts per second in HBase is $17$ times more in comparison to MySQL. The results are shown in Fig. 5(a) and Table 9. As can be seen from Fig. 4(b) and Fig. 5(a), the performance of HBase is better as compared to MySQL. For batch insertion MySQL uses InnoDB default buffer size of $8$ MB to add batch records in it until buffer reaches a threshold value or commit occurs. On the other hand, HBase stores all its batch records in HBase write client buffer which is configured as 20 MB in HBase configuration file. We perform an experiment with the same configuration. Thus, we change InnoDB buffer size from $8$ MB to $20$ MB.} 
\par{In HBase there is a lag in persisting the data stored on memstore to disk and it is by default asynchronous. On the other hand, MySQL persists data on disk and it is by default synchronous. To have the same configuration we change the durability of HBase to FSYNC\textunderscore WAL in HBase configuration file. FSYNC\textunderscore WAL writes the data to WAL synchronously and forces it to the disk. From the results it can be seen that time taken in HBase is $25$ times lower in loading $5,00,000$ records with different batch sizes as compared to MySQL. We believe the reason for this can be the difference in the way records are inserted in MySQL and HBase. In MySQL, executing an insert statement is a five step process. The batched insert statements in a buffer are first sent to the server, then parsed, then values are checked for uniqueness (intent hidden query), then data is inserted in actual table and finally data is inserted in index table. In HBase executing an insert statement is a two step process. The first step is writing the data to WAL then to the memstore and finally to the disk synchronously. Thus, the performance of HBase is better as compared to MySQL for batch insertion.}
\par{We also conduct a single row insertion experiment to examine which database can perform better for single row insertion. Fig. 5(b) and Table 10 reveal that the performance of HBase is better as compared to MySQL for all the datasets. The reason is same as batch insertion but here instead of sending records in a batch we are sending a single record in a single round trip. Fig. 5(b) reveals that when the dataset size increases then the difference between the time taken in loading real time data in MySQL and HBase also increases. We examine that the difference is $14$ times lower in HBase as compared to MySQL. Hence, performance of HBase is better as compared to MySQL in loading different datasets with single record insertion.}

\section{Conclusion}
\par{In this paper, we present an implementation of $\alpha$-miner algorithm in MySQL and HBase using SQL. Furthermore, we present the performance benchmarking and comparison of $\alpha$-miner algorithm in MySQL and HBase. The $\alpha$-miner implementation in MySQL is a one tier application which uses only standard SQL queries and advanced stored procedures. Similarly, implementation in HBase is done using Phoenix. We conclude that HBase on an average is $29$ times faster than MySQL in loading large datasets. Based on experimental results, we conclude that HBase outperforms MySQL in loading real time data (event logs) by having $17$ times more number of inserts per second.}
\par{We conclude the total time taken to read the data while execution of $\alpha$-miner algorithm is $1.16$ times lower in HBase as compared to MySQL. Similarly, for writing the data, time taken by HBase is $1.70$ times lower as compared to MySQL. We conclude the total execution time of $\alpha$-miner algorithm improves significantly in HBase as compared to MySQL by $1.46$x order of magnitude. HBase outperforms MySQL in terms of the disk usage of tables. The disk space occupied by tables in HBase is $4.37$ times lower as compared to MySQL. Thus, we conclude that HBase is more efficient than MySQL in terms of storing data and performing query. Using well known compression technique, HBase outperforms MySQL in disk usage as well as execution of $\alpha$-miner algorithm.} 

\nocite{*} 
\bibliographystyle{sql}
\bibliography{citation}
\end{document}